\documentclass[iop]{emulateapj}
\usepackage{threeparttable}
\usepackage{graphicx}
\usepackage{graphics}
\usepackage{color}           
\usepackage{url}             

\begin{document}

\title{Observations of an Energetically Isolated Quiet Sun Transient: Evidence of Quasi-Steady Coronal Heating}


%
\author{N. Brice Orange$^{1,2}$, David L. Chesny$^{1,3}$, and Hakeem M. Oluseyi$^{3}$}
\affil{$^1$OrangeWave Innovative Science, LLC, Moncks Corner, SC 29461}
\affil{$^2$Etelman Observatory, St. Thomas, United States Virgin Islands 00802}
\affil{$^3$Department of Physics \& Space Sciences, Florida Institute of Technology, Melbourne, FL  32901}

\begin{abstract}
Increasing evidence for coronal heating contributions from cooler solar atmospheric layers, notably quiet Sun (QS) conditions, challenges standard solar atmospheric descriptions of bright transition region (TR) emission. As such, questions to the role of dynamic QS transients in contributing to the total coronal energy budget are elevated. Using observations from the {\it Atmospheric Imaging Assembly} and {\it Heliosemic Magnetic Imager} on board the {\it Solar Dynamics Observatory}, and numerical model extrapolations of coronal magnetic fields, we investigate a dynamic QS transient energetically isolated to the TR and extruding from a common footpoint shared with two heated loop arcades. A non-casual relationship is established between episodic heating of the QS transient and wide-spread magnetic field re-organization events, while evidence is found favoring a magnetic topology typical of eruptive processes. Quasi-steady interchange reconnection events are implicated as a source of the transient's visibly bright radiative signature. We consider the QS transient's temporally stable ($\approx$\,35\,min) radiative nature occurs as a result of the large-scale magnetic field geometries of the QS and/or relatively quiet nature of the magnetic photosphere, which possibly act to inhibit energetic buildup processes required to initiate a catastrophic eruption phase. This work provides insight to the QS's thermodynamic and magnetic relation to eruptive processes quasi-steadily heating a small-scale dynamic and TR transient. This work elevates arguments of non-negligible coronal heating contributions from cool atmospheric layers in QS conditions, and increases evidence for solar wind mass feeding of dynamic transients therein.
\end{abstract}
%

\section{Introduction}
\label{sec:intro}

Solar atmospheric activity has been directly tied to the underlying photospheric magnetic field evolution, which exhibits a wealth of complexity across broad spatial and temporal scales \citep[{\it e.g.},][]{Dowdyetal1986SoPh,Schrijveretal1992AA,Pevtsovetal2003ApJ,UritskyDavila2012ApJ}. Coronal magnetic field complexities are driven by enhanced magnetic energy and helicity build up \citep{Klimchuk2006SoPh,UritskyDavila2012ApJ}; processes known to drive eruptive events like coronal mass ejections (CMEs), jets, and flares \citep[{\it e.g.},][]{Solankietal2006RPPh,Antiochosetal2007ApJ}. However, the origin and physical relation of magnetic field complexities to observed multi-thermal solar atmospheric plasma responses remains a fundamental problem in solar physics.


The fine-structure and dynamics of transient solar atmospheric phenomena ({\it e.g.}, bright points, flares, jets, etc.) occurring across broad environmental conditions, {\it i.e.}, active regions (ARs) to coronal holes (CHs), and temporal scales provide a direct utility for deciphering energy balance process(es) responsible for energizing and sustaining the hot outer solar atmosphere \citep{Zacharias2011AA,ArchontisHansteen2014ApJ}. Transient event observations indicate multi-thermal plasma compositions, as they are most often visible in multi-wavelength radative imagery
({\it e.g.}, \citealt{Canfieldetal1996ApJ,Madjarskaetal2003AA,Leeetal2011ApJ,Orangeetal2013SoPh272O}), and have revealed correlations with eruptive behavior, {\it i.e.}, bright points (BPs) with jets \citep{Madjarskaetal2012AA,Orangeetal2013SoPh272O}. The current consensus is, transient events are self-consistent with flare models, where magnetic reconnection serves as the central engine ({\it e.g.}, \citealt{Heyvaertsetal1977ApJ,ForbesPriest1984SoPh}), but the nature of the magnetic environments and evolutions therein responsible for their formation remains outstanding ({\it e.g.}, \citealt{ArchontisHansteen2014ApJ}).
\begin{figure*}[!t]
\begin{center}
 \includegraphics[scale=0.31]{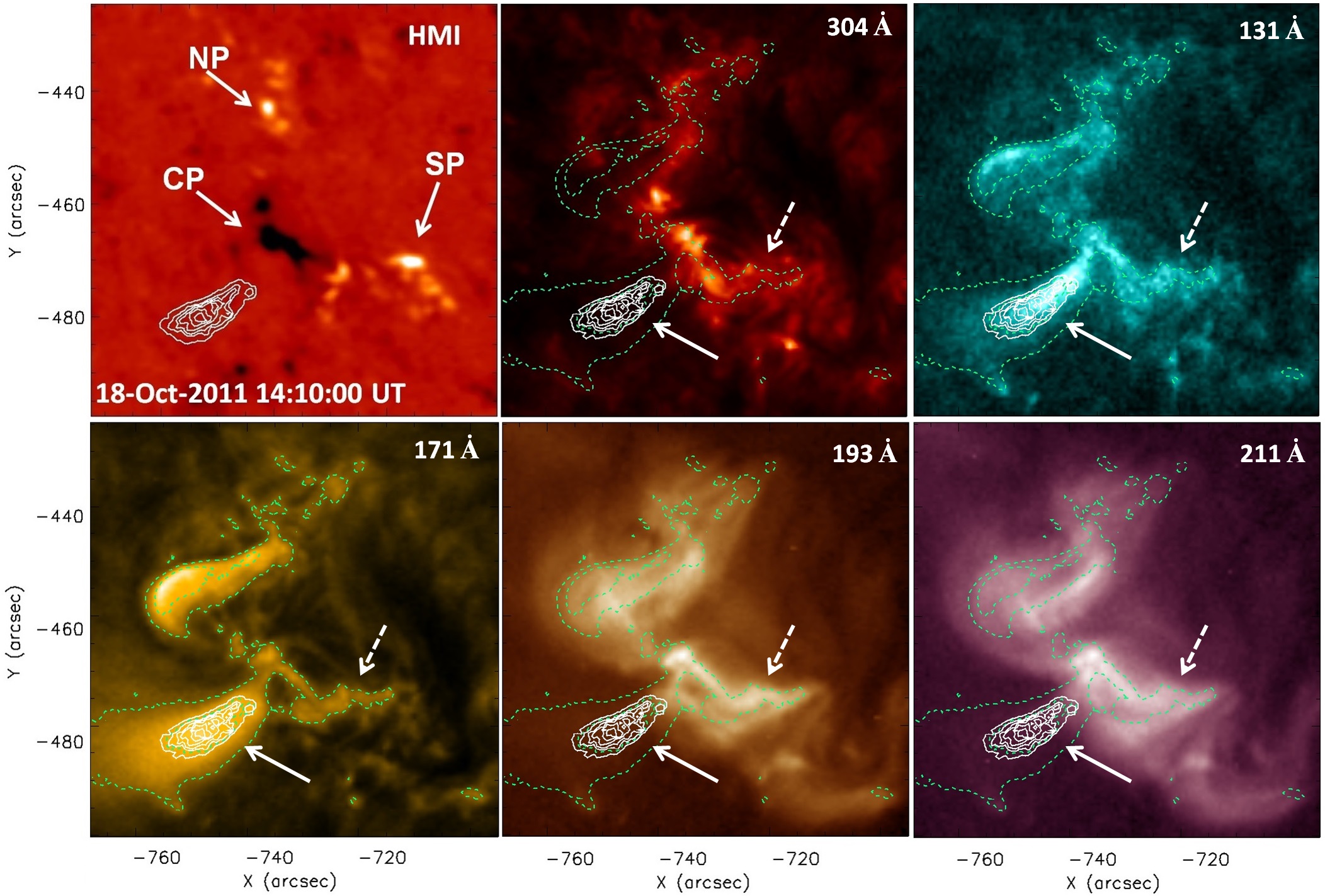}
 \caption{HMI LOS magnetogram and AIA radiative images, {\it i.e.}, 304\,{\AA}, 131\,{\AA}, 211\,{\AA}, 193\,{\AA}, and 171\,{\AA}, from top right to bottom right in clockwise fashion, respectively, observed 18 October 2011 at 14:10 UT. Solid blue contours (also identified by a solid arrow) highlight the QS feature, visually bright at only TR temperatures, and distributed away from a common footpoint shared with two heated loop arcades. Note, the dashed arrow highlights one of these hot coronal loop arcades, while the CP, NP, and SP sites (as discussed in the text) have been identified on the HMI image.}
\label{fig:SampleImageAlign_CoronalChromo_Evaporation}
\end{center}
\end{figure*}

In relation to proposed self-similar generation mechanisms for transient events, evidence to energetically isolated counterparts ({\it i.e.}, transition region BPs; \citealt{Orangeetal2013SoPh272O}) are of distinct interest, given their presence elevates questions to the role of varying thermodynamic conditions on the believed central engine
{\it e.g.}, possible variations in the efficiency of magnetic reconnection \citep{Longcope1998ApJ,LongcopeKankelborg1999ApJ}. Additionally, physical processes leading to the formation of dynamic transients with eruptive behavior, such as jets, are uniquely valuable. An inferred fundamental difference exists between jets and flares, as jets most commonly involve interchange reconnection between adjacent open and closed magnetic flux tubes ({\it e.g.}, \citealt{Shibataetal1992PASJ,WangSheely1993ApJ,Wangetal1996Sci,Shibataetal1997ASPC}), and is considered responsible for their generation.
Therefore, much stands to be learned from studies elucidating the radiative and magnetic field environments of energetically isolated transients, and those with eruptive behaviors. Particularly, deciphering the magnetic field's role in eruptive and non-eruptive processes in the presence of small-scale dynamic solar atmospheric transients.

\begin{table}[!t]
\begin{center}
\caption{AIA QS Transient Event Observations}
\label{tbl:EventLoc}
\vspace{0.5mm}
\begin{tabular}{c c c}
\hline \hline
Date & Time (UT) & Solar Coordinates (X, Y)  \\
\hline
18 October 2011 & 14:05:00 & -720, -455 \\
\hline
\end{tabular}
\end{center}
\end{table}

Jets, most abundantly observed in CHs
({\it e.g.}, \citealt{Shimojoetal1998SoPh,Savchevaetal2007PASJ}) and ARs ({\it e.g.}, \citealt{ShimojoShibata2000AdSpR,Schmiederetal2013AA}), are most commonly linked to reconnection events between emerging and pre-existing fields ({\it e.g.}, \citealt{YokoyamaShibata1995Natur}), where open magnetic fields are the key element in their collimated, and evaporating, plasma flows ({\it e.g.}, \citealt{Moreno-Insertisetal2008ApJ,Liuetal2009ApJ}).
To a lesser extent, they occur in regions of quiet Sun (QS), but tend to be characterized by slower propagation speeds and shorter collimated outflows \citep{TateArbacheretal2015AAS} in contrast to those of CHs and ARs, as well as correlate more often with BPs \citep{Orangeetal2013SoPh272O,TateArbacheretal2015AAS}.
\cite{YokoyamaShibata1996PASJ} noted, some QS jets reflected loop brightenings due to a pair of opposing horizontal flows that resulted from the general geometry of the large scale coronal magnetic fields, {\it i.e.}, a larger horizontal component than that of CHs \citep{Shibataetal1994xspy}.
Similarly, \cite{Orangeetal2013SoPh272O} reported, in a BP comparison study, a more horizontally directed current sheet ({\it i.e.}, relative to the solar surface normal) explained well the formation of a transition region (TR) energetically isolated QS BP, as a manifestation of self-similar plasma heating to its hotter coronal BP counterparts. The existence of evidence for the QS's capability of sustaining fine-scale structuring conducive to eruptive activity, and coronal heating \citep{Chesnyetal2013ApJL}, highlights the importance of identifying the fundamental processes leading to differing occurrence rates, and general characteristics of eruptive events in CHs and ARs versus that of the QS.

In light of the previous discussions, we have on hand a data set, derived from observations taken by the {\it Atmospheric Imaging Assembly} (AIA; \citeauthor{Lemenetal2012SolPh} \citeyear{Lemenetal2012SolPh}) and {\it Heliosemic Magnetic Imager} (HMI; \citeauthor{Schouetal2012SoPh} \citeyear{Schouetal2012SoPh}) on board the {\it Solar Dynamics Observatory} (SDO), that provide direct observational evidence of a
previously unexplored, and possibly ubiquitous QS phenomena. Our QS transient event was energetically isolated to TR temperatures, and distributed away from two heated loop arcades to which it shared a common footpoint (Figure~\ref{fig:SampleImageAlign_CoronalChromo_Evaporation}). Temporally analyzed radiative and LOS magnetic field imagery of the QS transient, and immediately surrounding regions, suggested its visibly bright radiative signature, derived from plasma heating {\it via} interchange reconnection events in an environment conducive to eruptive activity. Coronal magnetic field extrapolations are used as support for topologies that typically precede eruptive phenomena, {\it i.e.}, closed and open magnetic field interface. Based on the feature's temporally radiatively stable nature ($\approx$\,35\,min) physical sources are discussed that likely resulted in quasi-steady solar atmospheric energy re-distribution, over the large scale eruptions which typically accompany such magnetic topologies.

\section{Observations and Analysis}

Observational data was obtained from SDO's AIA and HMI instruments on 18 October 2011 from 14:05 UT -- 14:40 UT. The AIA data consists of the following passbands: 131\,{\AA} ($\log T$\,$\approx$\,5.8), 171\,{\AA} ($\log T$\,$\approx$\,5.9), 193\,{\AA} ($\log T$\,$\approx$\,6.2), 211\,{\AA} ($\log T$\,$\approx$\,6.3), and 304\,{\AA} ($\log T$\,$\approx$\,4.8). It is recognized, AIA passbands are multi-thermal; however, in QS conditions the dominant emission observed in each passband originates from those reported above \citep{ODwyeretal2010AA,Lemenetal2012SolPh}. In support of such notions, we found that the typical flux rates associated with our observations correlated with those measured from a large scale QS region, close to Sun center, observed the same day.

Each AIA passband images the Sun's full disk approximately every 12 s, and provides coverage of solar plasma from chromospheric to coronal temperatures with a spatial resolution of $\approx$\,0\farcs6. The HMI data are images of the full disk line-of-sight (LOS) magnetograms with a cadence of 45 seconds and spatial resolution of $\approx$\,0\farcs5. AIA passband and HMI LOS magnetograms were pre-processed using standard {\sf Solar SoftWare} (SSW), with pointing corrections of \citet{Orangeetal2013SoPhb}. Passband images were aligned at each time stamp via the SSW routine {\sf drot\_map.pro} and visual correlation of bright coronal structures with strong photospheric magnetic elements, with a resultant uncertainty of $\approx$\,0\farcs6 (Figure~\ref{fig:SampleImageAlign_CoronalChromo_Evaporation}).

Recall, observational data was selected due to its witnessing of a visually bright QS feature at only TR temperatures (Figure~\ref{fig:SampleImageAlign_CoronalChromo_Evaporation}). In relation to co-spatial underlying LOS magnetograms, this feature was distributed away from the the cusp of a horseshoe like feature of photospheric magnetic elements, characterized by order of magnitude larger field strengths than those of the surrounding background. The horseshoe magnetic feature, a higher order field configuration, was comprised of one strong negative and two weaker positive polarity elements. Note, the positive polarity elements lie in the positive solar $x$ direction, relative to the negative element of the cusp position. In the remainder of this document, the cusp position of the magnetic feature is referred to as the CP site, and the positive elements are referred to as the north and south positions, {\it i.e.}, NP and SP sites, respectively ({\it e.g.}, see Figure~\ref{fig:SampleImageAlign_CoronalChromo_Evaporation}).

It is the goal of the remainder of this manuscript to investigate the processes supplying the QS transient feature with heated plasma, and maintaining its visually bright radiative signature over an $\approx$\,35 min period. Similarly, the role of the underlying magnetic field and two hot coronal loop arcades are probed as they relate to the QS transient feature. We describe our methodology for investigating temporal variations of radiative and magnetic characteristics in $\S$~\ref{sec:RadMag_Env}. Techniques employed to investigate the associated 3D coronal magnetic fields in the features vicinity are described in $\S$~\ref{sec:ChpERPTR_MCME_MagFluxDens_CMSModel_Analysis}.
\begin{figure*}[!t]
\begin{center}
 \includegraphics[scale=0.23]{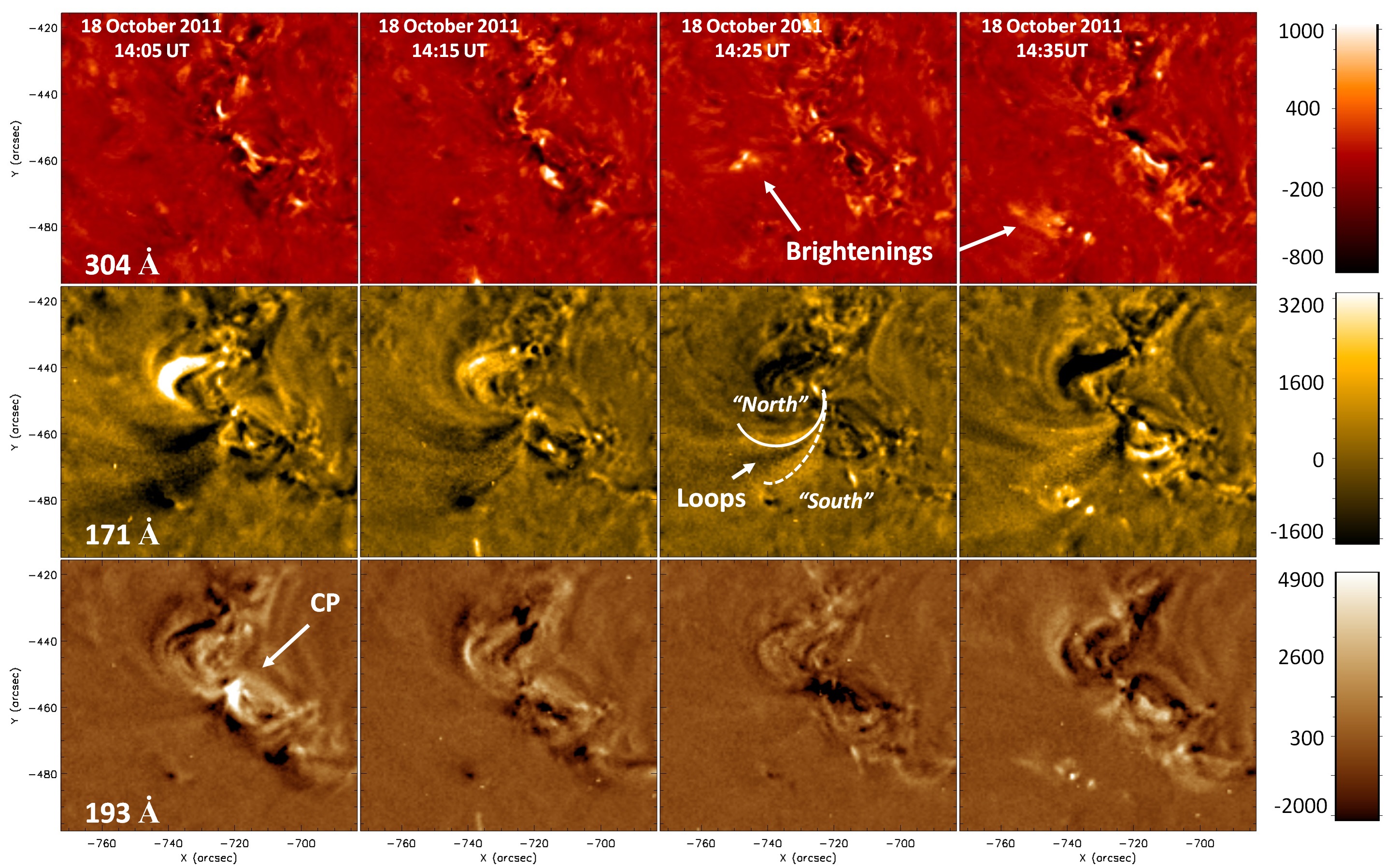}
 \caption{304\,{\AA}, 171\,{\AA}, and 193\,{\AA} images (DN pix$^{-1}$ s$^{-1}$), less image averages, from top to bottom, respectively, observed on 18 October 2011 covering 14:05 UT\,--\,14:35 UT at approximately ten minute intervals, from left to right, respectively. In these images we have highlighted the CP site, as well as some physically interesting sub-structure of the QS transient feature revealed during this process. Mainly, the appearance of ``loop-like" structures and small-scale brightenings.}
\label{fig:RDImage_Seq}
\end{center}
\end{figure*}

\subsection{Radiative and Photospheric Magnetic Environments}\label{sec:RadMag_Env}

The QS transient feature was characterized by a visibly bright radiative signature throughout our observational time domain, confined to AIA passbands dominated by emission lines formed at TR temperature regimes ({\it i.e.}, 171\,{\AA}). Here we point out, throughout the remainder of our manuscript the terminology of chromospheric, TR, and coronal temperature regimes are used interchangeably for 304\,{\AA}, 171\,{\AA} and 193\,{\AA} observations, respectively. To highlight sub-structure of our QS transient feature, we first obtained an average image, per AIA passband, over the entire duration of our presented observation sequence ({\it i.e.}, 14:05 UT\,--\,14:40 UT). Subsequently, this average image (baseline) was then subtracted off from all imagery, per respective passband. In that respect, our methodology utilizes radiative images, less their respective image average (covering the $\approx$\,40 min time domain), to highlight physically interesting sub-structure. In Figure~\ref{fig:RDImage_Seq} an example of this process has been provided. As revealed therein, the QS transient was comprised of ``loop-like sub-structures," e.g., see 14:25 UT in said figure. Visual inspection revealed these structures experienced multiple heating and cooling episodes ({\it i.e.}, brightenings and dimmings, respectively), while their footpoints, opposing those shared with the hot coronal loop arcades, brightened in chromospheric through coronal emission (Figure~\ref{fig:RDImage_Seq}). Note, for simplicity throughout the remainder of this manuscript, the aforementioned loop structures are referred to as the ``North" and ``South" loops, NL and SL, respectively.

In regards to the previously highlighted ``heating and cooling" episodes, it is recognized our use of AIA filtergrams, do not allow explicit determination if such intensity increases are a result of plasma heating or the presence of dense warm material \citep{Chandrashekharetal2014AA}. However, we emphasize, in relation to coronal heating it remains essential to consider it not only a process of increasing temperature, but density as well ({\it e.g.}, \citealt{Aschwandenetal2007ApJ,Uzdensky2007ApJ}). To that effect, throughout the remainder of our manuscript, episodes of increasing radiative emission are referred to as ``heating" events -- a coronal heating supply either directly through plasma heating or enhanced plasma densities.

\begin{figure*}[!t]
\begin{center}
 \includegraphics[scale=0.24]{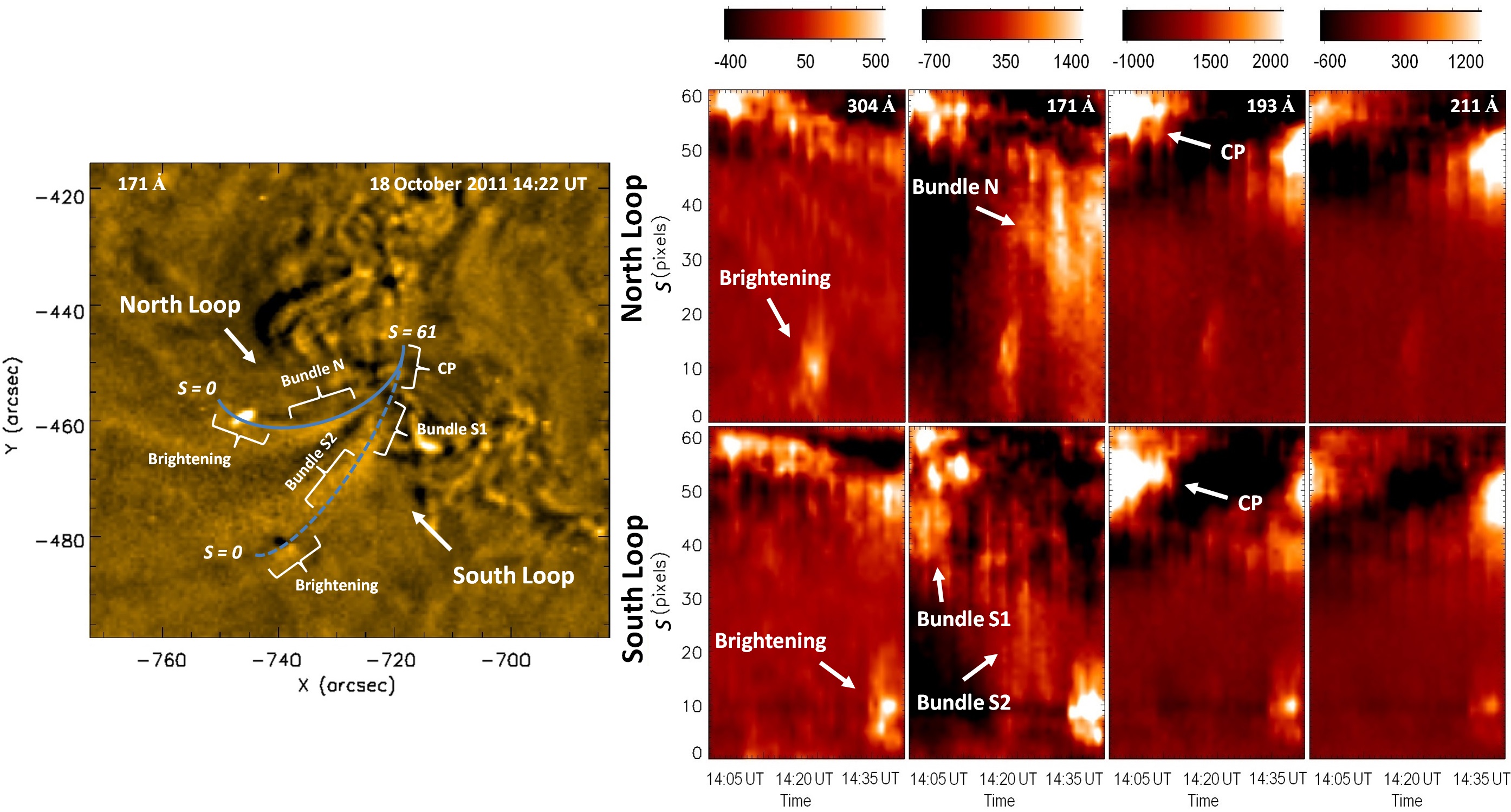}
 \caption{{\it Left:} 171\,{\AA} radiative image observed 18 October 2011 at 14:22 UT, with solid (blue) and dashed (blue) lines marking the spines of the NL and SL features, respectively. Additionally, denoted are the CP site, bundles, and brigthening segmented regions of each feature, as discussed in the text. {\it Right:} Radiative fluxes (DN pix$^{-1}$ s$^{-1}$), less image averages, smoothed as a function loop spine position, $s$, for the NL and SL features (top and bottom rows, respectively) versus time for 304\,{\AA}, 171\,{\AA}, 193\,{\AA}, and 211\,{\AA} observations on 18 October 2011 covering 14:05 UT\,--\,14:40 UT at approximately one minute intervals, from left to right, respectively. In these images we have also highlighted the CP site (same for both features), brightenings, and bundle regions indicative of localized heating for each feature.}
\label{fig:NS_Spires}
\end{center}
\end{figure*}

To investigate NL and SL radiative emission modulations we applied a semi-supervised tracing algorithm to 171\,{\AA} images that assigns a set of loop coordinates $s$ along their spines \citep{Orangeetal2013ApJ}. Loop tracings were then used to aggregate data from all remaining passbands, as well as segment loops into sub-regions of: their common footpoint shared with the hot coronal loop arcades (i.e., CP site), opposing footpoint regions, and localized regions of radiatively enhanced TR emission ({\it i.e.}, ``bundle" regions). Here we note, NL and SL segmentation were performed {\it via} a combination of visual inspection of image average subtracted observations and their cross-correlation to loop cross section smoothed radiative fluxes as function of $s$, with errors derived from photon counting statistics ({\it e.g.}, Figure~\ref{fig:NS_Spires}). Light curves for segmented regions were then generated by integrating the 3$\sigma$ brightest flux less the running average, and standard error prorogation techniques.

It was highlighted previously, we wish to study the possible role of the higher order underlying magnetic field configuration in our QS transient's heated plasma supply; particularly, this is the LOS magnetic fields of the NP, SP, and CP sites, identified above. Selection of these magnetogram regions, throughout our observational time-domain is detailed as follows. First, at the onset of our observation sequence (14:05 UT), the magnetogram, with 171\,{\AA} and 193\,{\AA} intensity contours overlaid ({\it e.g.}, similar to that presented in Figure~\ref{fig:SampleImageAlign_CoronalChromo_Evaporation}) was visually inspected, and utilized to define the three footpoint regions highlighted by the contoured intensity isolines. Then at each of the footpoint sites, a boxed region was defined which encompassed the corresponding photospheric magnetic element, as well as some surrounding background. Throughout the remainder of our observation sequence, at approximately one minute intervals, similar visual inspection techniques were employed, in order to minimize instances of false emergence, cancelation, or fragmentation events due to magnetic flux movement out of, or into the selected fields-of-view.

\begin{figure}[!t]
\begin{center}
 \includegraphics[scale=0.31]{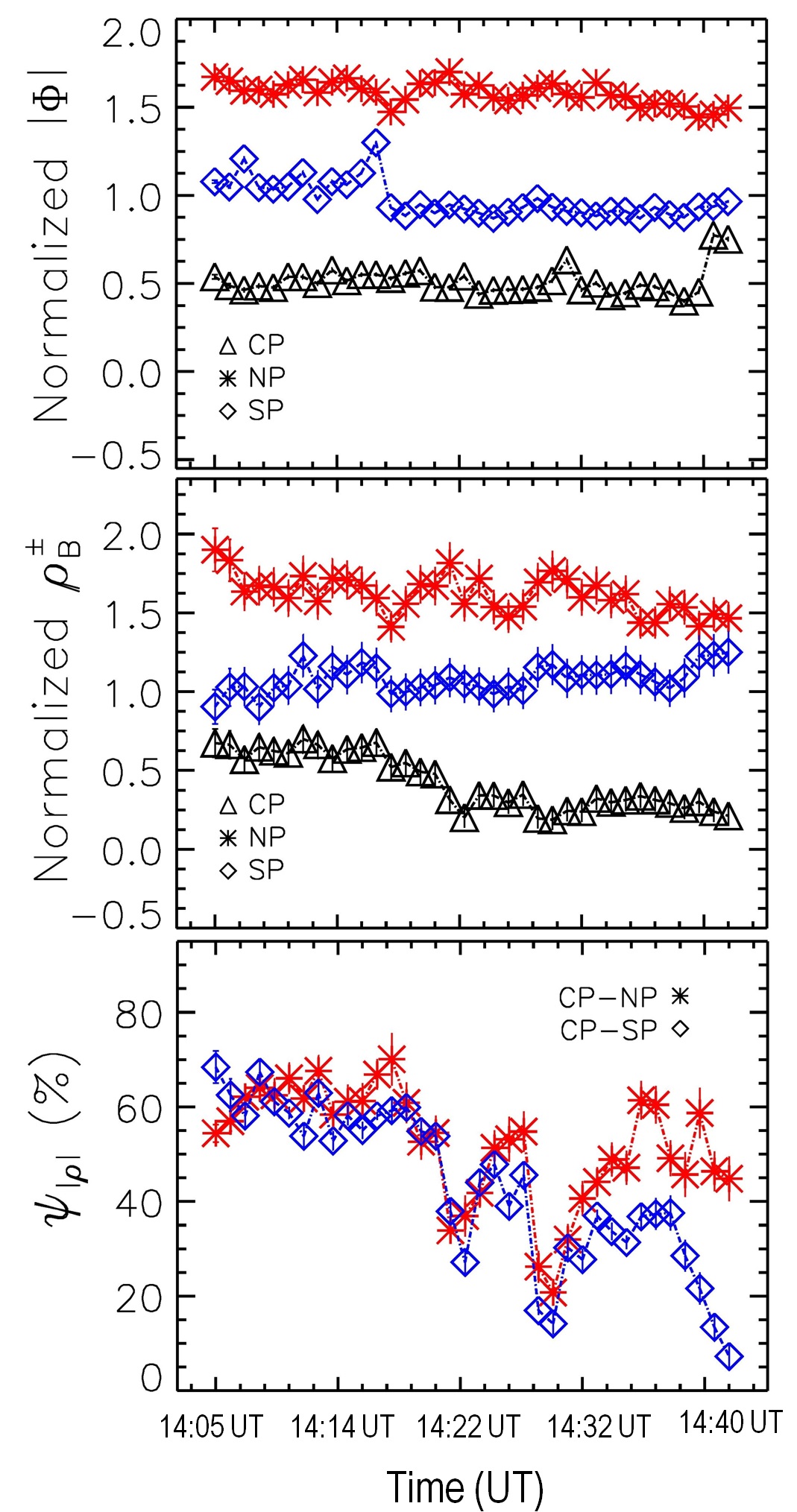}
 \caption{Unsigned magnetic flux (arbitrary units), flux densities (arbitrary units), and flux density gradients (\%) from top to bottom, respectively, observed on 18 October 2011 over 14:05 UT\,--\,14:40 UT. Note, in the unsigned magnetic flux and flux density plots the data have been scaled for visualization purposes.}
\label{fig:MC_for_Event}
\end{center}
\end{figure}

Temporal variations in the unsigned magnetic flux $|\Phi|$ were then measured from an integration of the positive and negative flux elements in the above described NP, SP, and CP regions, with errors propagated from photon counting statistics (Figure~\ref{fig:MC_for_Event}). Within each region, magnetic flux densities ($\rho^{\pm}_B$), defined as the number of positive and negative flux elements above and below a threshold of 20\,G \citep{DeForsetetal2007ApJ,UritskyDavila2012ApJ}, {\it i.e.},
\begin{equation}
\rho_{|B|} = \frac{\sum_N I(|B|)}{N},
\end{equation}
where $N$ is the number of pixels in the FOV, $|B|$ is the LOS magnetic field, and $I(|B|)$ is an indicator function defined as
\begin{equation}
I(|B|) = \Big \{
\begin{array}{cc}
1, & \hspace{.05in} |B| > {\rm 20\,G} \\
0, & \hspace{.05in} \textrm{Otherwise}, \\
\end{array}
\end{equation}
were additionally measured \citep{Orangeetal2013ApJ,Chesnyetal2013ApJL}. Note, these results allow investigations on flux likely contributing to reconnection events \citep{Sakaietal1997SPD,Orangeetal2013ApJ,Chesnyetal2013ApJL}. Additionally, magnetic flux densities were used as a means for estimating photospheric magnetic flux imbalances, $\psi_{|\rho|}$, relative to the CP from the NP and SP sites, {\it i.e.},
\begin{equation}\label{eqn:MagFluxImb}
\psi^{{\rm NP,SP}}_{|\rho|} = \frac{\rho^{{\rm NP,SP}}_{B} - \rho^{{\rm CP}}_{B}}{\rho^{{\rm CP}}_{B}}.
\end{equation}
Uncertainties in both magnetic flux densities, and their respective gradients were derived from a combination of photon counting statistics and standard error propagation techniques. We recognize there is not a real flux imbalance, and that such measurements are strongly influenced by both the region from which the measurement is derived, as well as those below our applied threshold. As such, we emphasize our $\psi_{|\rho|}$ measurements are only used as a qualitative assessment of the general magnetic system, and fluxes therein likely contributing to reconnection processes, where our QS transient was witnessed.

\subsection{Coronal Magnetic Field Environment}\label{sec:ChpERPTR_MCME_MagFluxDens_CMSModel_Analysis}

The coronal magnetic field environment of our QS transient was investigated by employing the Coronal Modeling Software (CMS; van Ballegooijen 2011, private communication). CMS, written in the {\sf Interactive Data Language} (IDL) and {\sf Fortran90}, describes the 3D coronal magnetic structure as a function of a single instant in time using the non-linear force-free method, {\it e.g.}, \citealt{VanBalletal2000ApJ539,MackayVanBall2006ApJ641}, to extrapolate coronal magnetic fields from observations of the LOS magnetic photosphere. CMS deduced coronal magnetic field models rely on a HMI LOS magnetogram, serving as a lower boundary condition, and a Synoptic Optical Long-Term Investigations (SOLIS; \citealt{Kelleretal1998ASPC}) Carrington rotation map describing the full solar disk photospheric magnetic field for computation and construction of a global potential field model. Each resultant empirical field model is comprised of a high and lower resolution area, which both depend on spherical coordinate systems, {\it i.e.}, $r$, $\theta$, $\phi$, where $r$ is the radial distance from the Sun's center, $\theta$ is the angle relative to the axis of solar rotation, and $\phi$ is the azimuthal angle.

\begin{figure}[!t]
\begin{center}
 \includegraphics[scale=0.3]{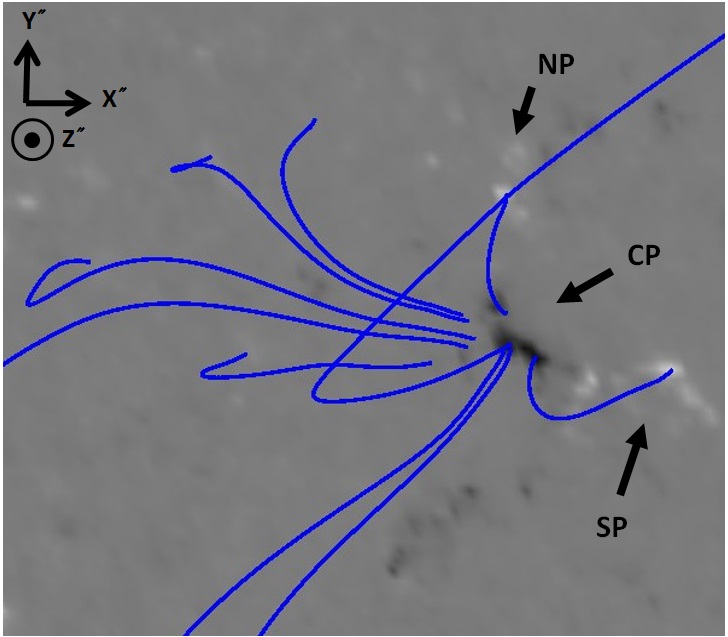}
 \caption{3D CMS model of the coronal magnetic fields related to our QS transient feature, correlating with the 18 October 2011 14:15 UT LOS HMI magnetogram observation, and where the CP, NP, and SP sites have been identified therein.
 }
\label{fig:CMSIntro_Image}
\end{center}
\end{figure}

The high-resolution (HIRES) region represents a discrete volume, large enough such that accounting for spherical geometry is required. In our case, the HIRES volume reflects a lower boundary area defined by a HMI LOS magnetogram extending to a height, $Z$, at which the field lines become radial, {\it i.e.}, $Z$ $>$ $R_{\sun}$. Potential magnetic fields of the HIRES regions are computed using spherical harmonics and eigenmode solutions, where the HIRES boundary ({\it i.e.}, that with the lower resolution region) is satisfied by a continuous normal magnetic field component everywhere. Everywhere in the HIRES volume, the magnetic field is defined by,
\begin{equation}
{\bf B} = \nabla \times {\bf A},
\end{equation}
with A the vector potential field, such that,
\begin{equation}
\nabla \cdot {\bf B} = 0,
\end{equation}
is satisfied, while initial non-potential field conditions are setup from flux rope insertion to potential fields and then subsequently relaxed using magnetofrictional relaxation techniques. Note, variable grid spacing is used within this region to make possible construction of models which extend to large coronal heights. The lower resolution region, referred to as the global region, is represented by a current-free potential field, derived using uniform grid spacing and the synoptic map of the radial magnetic field in the photosphere.

The radial component of the magnetic field, $B_r$, in the HIRES and global regions are defined initially by the lower boundary photospheric observations of the magnetograms and synoptic maps, respectively. Then, the potential field ${\bf B}({\bf r})$ associated with the observed photospheric flux distribution, $\tilde{B}_r (\theta,\lambda)$ at $r = R_\odot$ in spherical geometry, is calculated.

Note, CMS coronal magnetic field models then allow field lines to be traced, ``visualized," interactively by selecting specific flux elements in the photospheric field of view for which modeled results should be presented ({\it i.e.}, field lines originating at that footpoint site). Here we emphasize, our interest in CMS generated 3D coronal magnetic field models lies in a qualitative assessment of the general field environment of our QS transient event. Thus, we restrict field model assessments to those supportive of our observations, {e.g.}, Figure~\ref{fig:CMSIntro_Image}. Here we emphasize, as observed in Figure~\ref{fig:CMSIntro_Image}, it supports previous indications that the QS transient was comprised of closed field structures ($\S$~\ref{sec:RadMag_Env}), and further suggests they are in close proximity to open fields.

\begin{table*}[!t]
\begin{center}
\caption{Event Milestones}
\label{tbl:EventMileSt}
\vspace{0.5mm}
\begin{tabular}{c c }
\hline
Time & Milestone \\
\hline
&  \\
-- {\bf Episode \# 1} -- & \\
&  \\
14:05\,--\,14:07 &  Magnetic cancelation at CP and NP sites, emergence at SP \\
14:06\,--\,14:09 &  Enhanced emission of CP and NL and SL bundles, which lags to cooler layers \\
14:11 &  SL loop ruptures - visually bright TR upper and lower bundle, later bright in coronal regime; \\
& SL opposing footpoint brightens \\
14:12 &  Upper TR bundle laterally drifts south, while both upper and lower bundles fade \\
14:13\,--\,14:14 &  SL opposing footpoint brightening peak, then completely faded \\
 & \\
-- {\bf Episode \# 2} -- & \\
&  \\
14:26\,--\,14:28 & Magnetic cancelation at CP site, SP and NP emergence \\
14:28\,--\,14:30 & Enhanced emission of CP, and NL, and SL bundles, which lags to cooler layers \\
14:30 & Spire-like structure in NL vicinity, more extended in TR \\
14:31\,--\,14:32 & Lateral expansion of TR spire, and evolving towards a more bundled appearance \\
14:32\,--\,14:33 & Dimming of bundled plasma as it drifts north and outward (away from CP) \\
14:34 & NL and SL opposing footpoints radiative flux increases; latter only visually bright \\
& \\
-- {\bf Episode \# 3} -- & \\
& \\
14:30\,--\,14:34 & Magnetic cancelation at CP site, and NP emergence \\
14:35\,--\,14:40 & Enhanced emission of CP, and NL, and SL bundles; \\
                 & SL opposing footpoint peaks in chromosphere \\
14:37 & Multiple spire-like structures in TR, mainly in vicinity of NL, and distinct (shorter) coronal counterpart  \\
14:38 & Shrinking of spire structures in TR; SL opposing footpoint peaks in TR and corona \\
14:39\,--\,14:40 & Enhanced emission of CP, NL, and SL, again with TR spire-like structures discernable \\
 & \\
\hline
\end{tabular}
\end{center}
\end{table*}

\section{Results}\label{sec:Results}

\begin{figure*}[!t]
\begin{center}
 \includegraphics[scale=0.19]{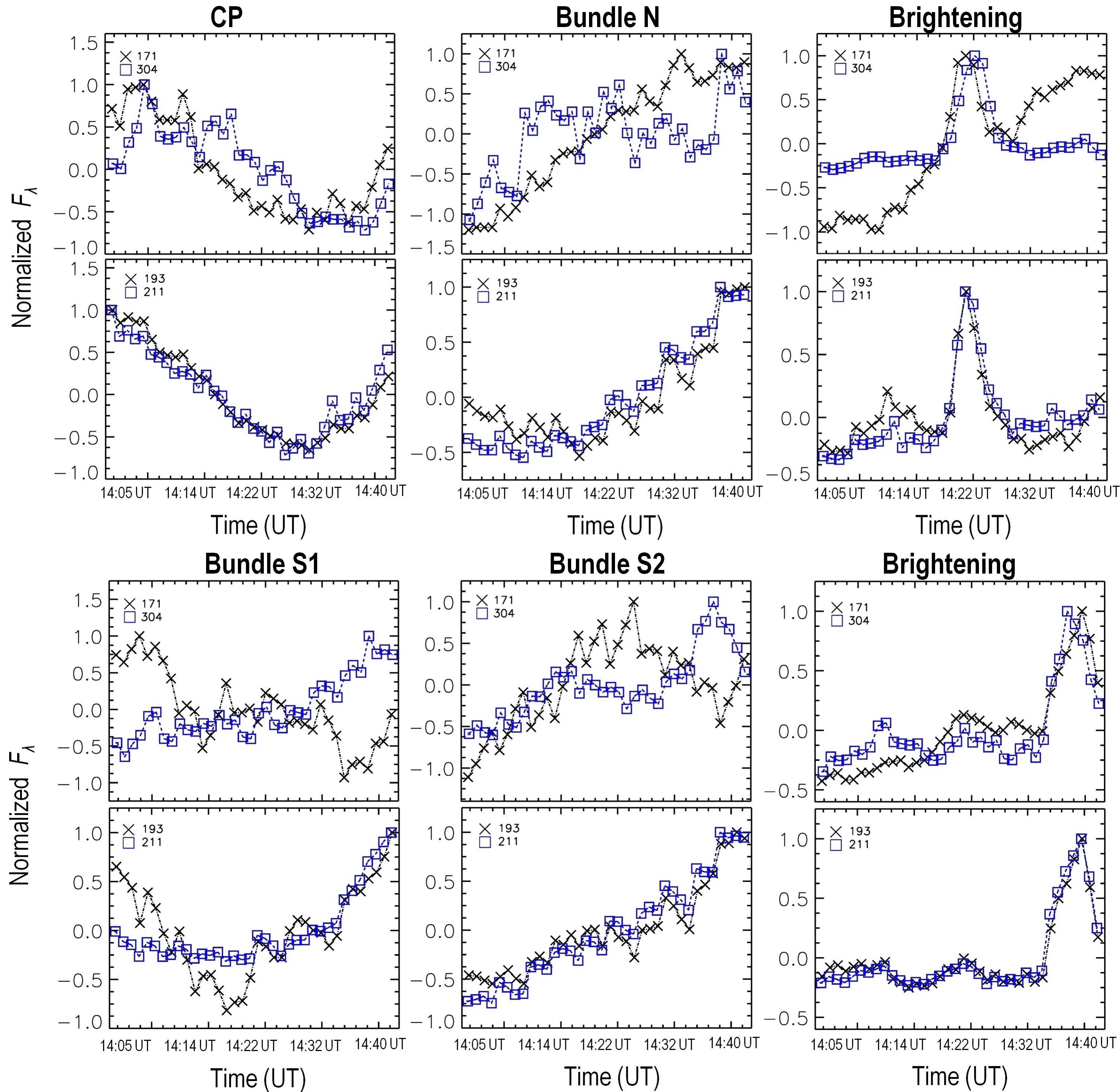}
 \caption{3$\sigma$ brightest fluxes (arbitrary units) of the NL and SL segmented regions (top and bottom panels, respectively) identified in Figure~\ref{fig:NS_Spires}, and labeled accordingly herein, versus time for chromospheric through coronal passbands.}
\label{fig:SubStr_NS-SS-LC}
\end{center}
\end{figure*}

Multi-wavelength AIA observations covering 14:05 UT\,--\,14:40 UT revealed the QS transient event maintained its visibly bright nature, confined to TR temperature regimes, consistent with that observed in Figure~\ref{fig:SampleImageAlign_CoronalChromo_Evaporation}.
Similarly, the two loop arcades, in which the QS transient feature shared a common footpoint, were visually bright in TR and coronal images during our observational time frame. When analyzed in radiative imagery less image averages, the QS transient feature provided evidence to localized episodes of heating and cooling; particularly, in relation to the NL and SL structures (Figure~\ref{fig:NS_Spires}). These localized ``sub-regions," of both the NL and SL structures, have been identified in Figure~\ref{fig:NS_Spires}, and their respective temporal radiative flux variations are presented in Figure~\ref{fig:SubStr_NS-SS-LC}, derived from techniques discussed in $\S$~\ref{sec:RadMag_Env}.

Investigations of the NL and SL sub-region radiative emission variations (Figure~\ref{fig:SubStr_NS-SS-LC}), and those of the underlying magnetic field (Figure~\ref{fig:MC_for_Event}), indicated the heating episodes exhibited a general similarity. This was, wide-spread magnetic field modulations at the NP, SP, and CP sites ({\it i.e.}, combination of emergence and cancelation), $\approx$\,2\,min\,--\,4\,min, preceded enhanced plasma emission of the CP site and some variation of the NL and SL sub-regions, $\approx$\,3\,min\,--\,4\,min. Then, either or both the NL and SL opposing footpoints brightened, which in some instances began in conjunction with the aforementioned heating events, $\approx$\,1 min\,--\,3 min. To that effect, the remainder of this section explicitly details the episodes during 14:05 UT\,--\,14:15 UT and 14:26 UT\,--\,14:40 UT, to shed light on the dynamic heating that is likely the source of the QS transient feature's continually bright radiative nature. Table~\ref{tbl:EventMileSt} lists event milestones witnessed during the aforementioned observational time-frames that will be reported on in detail below.

The onset of the ``first episode" (14:05 UT\,--\,14:07 UT) reported on here, was accompanied by an approximately equivalent magnetic flux density decrease of $\approx$\,15\,\%, at both the NP and CP sites, while that of the SP increased by $\approx$\,7\,\%. Similar trends of these sites unsigned magnetic fluxes accompanied the afore described flux density modulations, {\it i.e.}, coupling of decreasing unsigned magnetic flux and flux density and vice-versa (Figure~\ref{fig:MC_for_Event}). Magnetic flux density gradients of the three sites, considered here suggestive of the magnetic systems energy balance, dropped by $\approx$\,10\,\% between that of the CP SP, while the CP NP increased by $\approx$\,8\,\%.

The CP, Bundle S1 (BS1), and Bundle N (BN) regions experienced radiative enhancements as the wide-spread magnetic field modulation ceased (Figure~\ref{fig:SubStr_NS-SS-LC}). Enhanced plasma emission of cooler atmospheric layers, {\it i.e.}, the chromosphere and TR, were more pronounced at the CP site, than that of the corona, particularly, for increasingly hotter coronal temperatures. However, the peak radiative emission at these cooler layers did lag that of the corona by $\approx$\,2\,min. In relation to the localized NL event, chromospheric emission peaked prior to that of both the TR and corona. At the BS1 site, both coronal and TR enhancements preceded that of the chromosphere, whose subsequent peak occurred in conjunction with a second 171\,{\AA} emission increasing episode, not witnessed in coronal passbands (Figure~\ref{fig:SubStr_NS-SS-LC}).

\begin{figure*}[!t]
\begin{center}
 \includegraphics[scale=0.205]{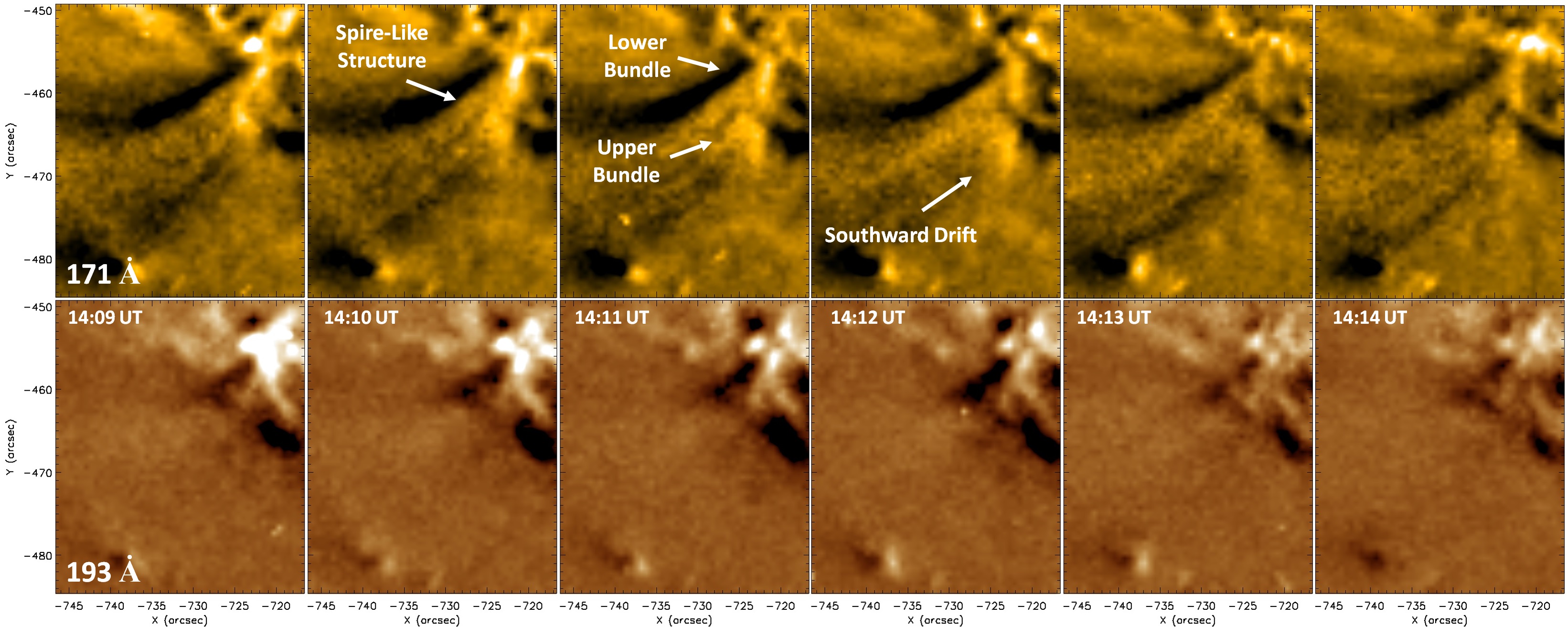}
 \caption{TR (171\,{\AA}) and coronal (193\,{\AA}), top and bottom rows, respectively, radiative images sequences, less image averages, observed 18 October 2011 during 14:11 UT\,--\,14:14 UT covering the observational time frame correlating with our QS transients heating episode \#\,1 as discussed in the text and outlined in Table~\ref{tbl:EventMileSt}. Note, herein, we have pointed out the appearance of a spire like structure, which approximately one minute later has appeared to rupture into two distinct plasma bundles, with the upper portion drifting southward and cooling at later times.}
\label{fig:SubStrSeq_E1}
\end{center}
\end{figure*}

In Figure~\ref{fig:SubStrSeq_E1}, the previously described localized heating of the SL is directly observed as a discernable spire-like structure, protruding from the CP site at $\approx$\,14:10 UT. As observed in accompanying 193\,{\AA} emission of this same figure, the 171\,{\AA} spire feature is extruding from a bright patch of coronal plasma, with a decreasing intensity distribution away from the CP site. By 14:11 UT, the spire-like structure has split, possibly evidence to an eruption \citep{Liuetal2011ApJ}, into two disjoint bright patches. The lower portion, remaining as a protrusion from the CP site, slowly faded thereafter. The upper portion, {\it i.e.}, that furthest from the CP site and only discernable in TR passbands, slowly fades, but additionally drifts outward, away from the possible eruption site, and southward. Interestingly, the upper portions complete disappearance ({\it i.e.}, 14:13 UT of Figure~\ref{fig:SubStrSeq_E1}) coincides with the radiative peak of the SL's opposing footpoint; visually discernable in chromospheric through coronal imagery. This feature then fades from view, first in chromospheric and coronal emission followed by that of the TR.

During the 14:26 UT\,--\,14:28 UT, the ``second episode" presented here, another wide-spread magnetic flux density modulation occurred (Figure~\ref{fig:MC_for_Event}). This instance witnessed a CP site flux density cancelation of $\approx$\,20\,\%, with emergence of $\approx$\,12\,\% and 9\,\% at NP and SP sites, respectively. Increased unsigned magnetic fluxes of the three sites accompanied the flux density modulations, however, considered negligible here given their marginal enhancements ($\lesssim$\,10\,G). Distinct decreases of magnetic flux density gradients correlated with this time interval (Figure~\ref{fig:MC_for_Event}); $\approx$\,30\,\% between both that of the CP SP, and CP NP sites, which is consistent with the reported flux density trends.

As the above magnetic field event concluded, emission increases of the CP, SL and NL bundled regions began (Figure~\ref{fig:SubStr_NS-SS-LC}). This CP site event lacked a chromospheric signature, while that of the BN region peaked first in the corona, and then lagged to cooler atmospheric layers. Note, the BS1 radiative emission enhancements immediately following the magnetic field re-organization event, was confined to TR and chromospheric regimes, but followed from an enhancement to coronal emission (only in 193\,{\AA} observations) that occurred in conjunction with the magnetic field event. In the BS2 region, radiative fluxes exhibited similar trends to that reported for the BN region, in that coronal emission peaked prior to that of cooler temperature regimes after the conclusion of the magnetic field re-organization event.

\begin{figure*}[!t]
\begin{center}
 \includegraphics[scale=0.205]{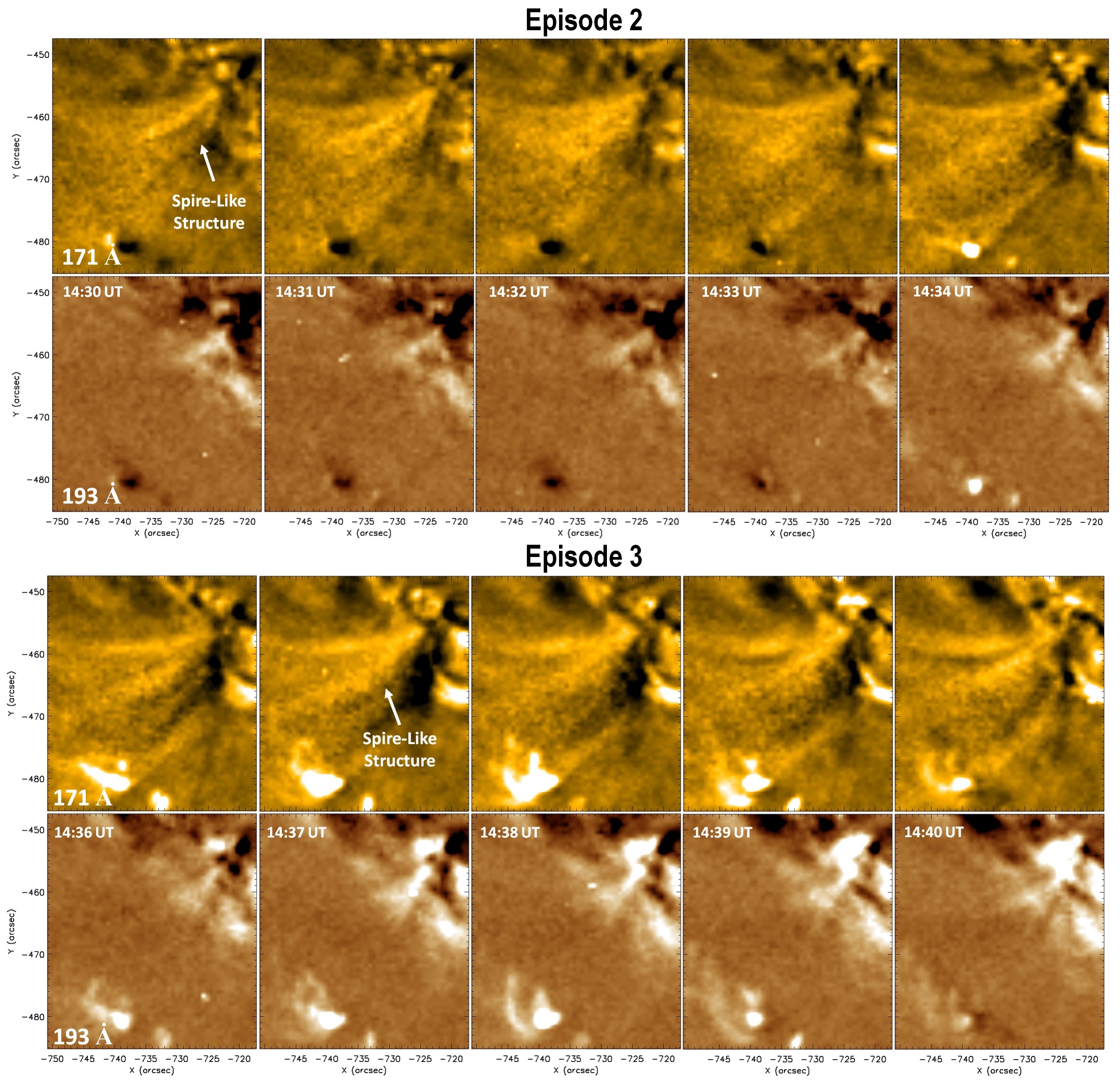}
 \caption{Radiative image sequences, presented similar to Figure~\ref{fig:SubStrSeq_E1}, but for our QS transients features heating episodes \#\,2 and \#\,3 (top and bottom panels, respectively) as as discussed in the text and outlined in Table~\ref{tbl:EventMileSt}. Note, in relation to each episode we have pointed out the appearance of a spire-like structure.}
\label{fig:SubStrSeq_E2}
\end{center}
\end{figure*}

We emphasize, NL and SL bundle regions peaked in coronal emission at $\approx$\,14:30 UT, and directly correlateda with the onset of another spire-like structure (Figure~\ref{fig:SubStrSeq_E2}). This feature was similar to that described for episode one (Figure~\ref{fig:SubStrSeq_E1}), but with a more northward location, and greater extent away from the CP site. Over the course of the next minute, this spire laterally expanded southward, as well as slightly increased in length. Additionally, as observed in Figure~\ref{fig:SubStrSeq_E2}, a possible second spire-like structure has emerged, slightly southward that of the first. At this time, we point out both spire-like structures, discernable at only TR regimes, suggest they are emanating from a bright, similar in structure but shorter, coronal feature. The southward feature additionally hints to another possible eruption, {\it i.e.}, a splitting into an upper and lower portion. Over 14:31 UT\,--\,14:33 UT, the previously bright TR emission of these structures, which is now more indicative of a single bright bundle of plasma and possibly a result of the previous evidence favoring an eruption, is observed to expand away from the CP site, and dim. Here we speculate the enhanced radiative emission in the vicinity of the NL, by 14:33 UT (Figure~\ref{fig:SubStrSeq_E2}), is likely evidence to said loop's apex \citep{Liuetal2011ApJ}. By 14:34 UT said upper TR plasma bundle has almost completely cooled, and given rise to enhanced emission of the CP site and SL opposing footpoint, the latter of which was visually bright in chromospheric through coronal passbands.

The remainder of our observation sequence, {\it i.e.}, 14:34 UT\,--\,14:40 UT, is again indicative of another heating episode. As consistent with the previous ones, this episode starts with a magnetic field re-organization characterized by CP magnetic flux density emergence ($\approx$\,12\,\%) and NP cancelation ($\approx$\,10\,\%), that is followed by wide-spread emission enhancements of the CP, NL, and SL localized regions. Thus, we detail below the observed plasma dynamics starting with the 14:36 UT observation time;
previously pointed out above in Figure~\ref{fig:SubStrSeq_E2}.

As observed at 14:36 UT in Figure~\ref{fig:SubStrSeq_E2}, discernable bright TR patches exist, in relation to our NL feature, and in a similar region as witnessed in episode two ({\it i.e.}, see the 14:32 UT observations of same figure). Note, the opposing SL footpoint has peaked radiatively in chromospheric emission at this time (Figure~\ref{fig:SubStr_NS-SS-LC}). Again, consistent with previous episodes, 14:37 UT observations witness the appearance of another spire-like structure that protrudes from the CP site, confined to TR regimes expect for an accompanying, shorter, coronal counterpart. It is also pointed out at this time, slightly north of the previously described feature, the NL region presents evidence to a similar, possibly even more, collimated structure (Figure~\ref{fig:SubStrSeq_E2}). Over the course of the following two minutes, the more southward spire-like structure of this episode has shrunk in length and dimmed, while the NL feature favors a similar evolution to that witnessed in each of the previous episodes. That is, the appearance of two distinct bright patches. In conjunction with these events, the SL's opposing footpoint radiatively peaks (Figures~\ref{fig:SubStr_NS-SS-LC} and \ref{fig:SubStrSeq_E2}), though this time in TR and hotter atmospheric layers. By the conclusion of our observational sequence, a number of sub-structures can be observed within our QS transient feature. To summarize, our 14:40 UT observations was characterized by enhanced radiative emission of various localized regions of the NL and SL features, and the re-emergence of a bundle of bright TR plasma, occurring in the region separating the two. These features again seemingly protrude from a common footpoint, visually bright in chromospheric through coronal emission, where coronal imagery supports a more compact spire-like structure.

\section{Discussion}\label{sec:Disc}

Investigations of temporal radiative emission across large scale solar atmospheric temperature gradients and co-spatial LOS magnetograms of our QS transient feature, as well as the CP site ({\it i.e.}, footpoint shared with two hot coronal loop arcades) provided evidence for plasma heating as a result of wide-spread magnetic field re-organization events. It was pointed out, {\it e.g.}, Figures~\ref{fig:RDImage_Seq} and \ref{fig:NS_Spires}, the QS transient revealed a dynamic ``sub-structure" composition. Detailed investigations ($\S$~\ref{sec:Results}), indicated the QS transient feature sub-structure was comprised of episodically heated (Figures~\ref{fig:NS_Spires} and \ref{fig:SubStr_NS-SS-LC}) open and closed magnetic fields. Of distinct interest was that episodic heating events exhibited a similar trend, which we summarize as follows.

Prior to each episode a re-organization of the underlying LOS magnetic environment of the QS transient feature ({\it i.e.}, our CP, NP, and SP sites), such as emergence and/or cancelations of magnetic flux densities, was observed (Figure~\ref{fig:MC_for_Event}). Enhanced radiative emission at various localized regions, established in conjunction with the QS transient feature ({\it i.e.,}, CP site, and/or NL and SL features), followed. Commonly the appearance of either a co-spatial, or adjacent spire-like feature, relative to either or both the NL or SL features ({\it e.g.}, Figure~\ref{fig:SubStrSeq_E2}), and/or the ``splitting" of heated loop structures into two disjoint patches of bright emission ({\it e.g.}, Figure~\ref{fig:SubStrSeq_E1}), accompanied the localized radiative enhancements. Heating episodes concluded with enhanced radiative emission events of the opposing footpoints of either or both the NL and SL features. In that respect, below we speculate to the source of the episodic heating, and its relation to our QS transient feature's unique visibly bright radiative signature. In compliment, discussions are presented to the general magnetic field environment and its temporal evolution, as it relates to our QS transient feature.

\begin{figure*}[!t]
\begin{center}
 \includegraphics[scale=0.305]{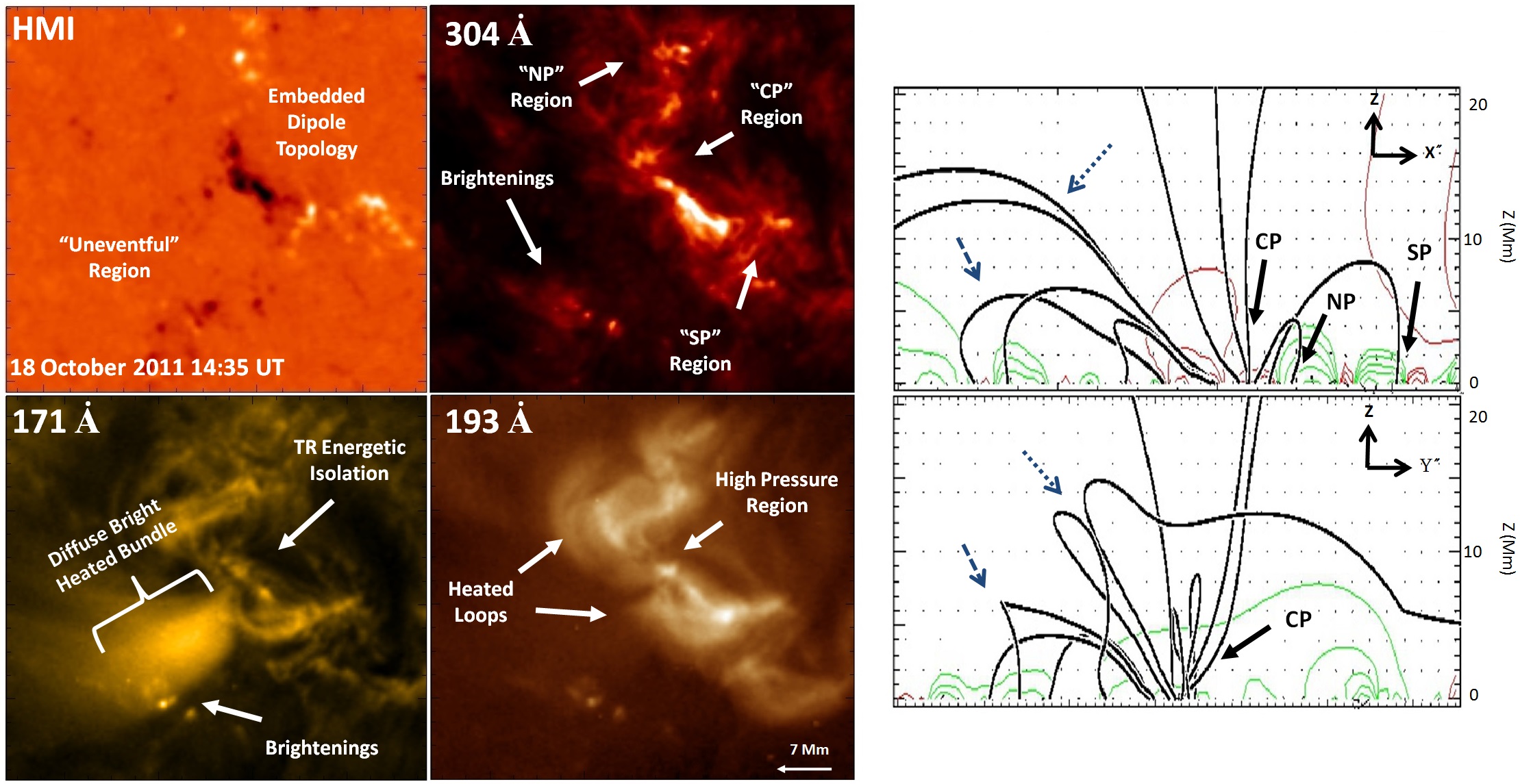}
 \caption{{\it Left:} HMI, 304\,{\AA}, 171\,{\AA}, and 193\,{\AA} images, from left to right and top to bottom, respectively. Throughout these images we have highlighted features extensively discussed in the text, and pointed out physically interesting features supportive to a magnetic environment common of eruptive processes. {\it Right:} 3D CMS X-Z and Y-Z (top and bottom, respectively) cross slices of the modeled coronal magnetic field image presented in Figure~\ref{fig:CMSIntro_Image}. On these images the CP, NP, and SP sites have been identified, and the dashed and dotted (blue) arrows point to field lines suggestive of closed and open magnetic fields in the vicinity of our QS transient feature.}
\label{fig:Disschematic}
\end{center}
\end{figure*}

Evidence for loop rupturing, and open magnetic fields are suggestive to interchange reconnection events ({\it e.g.}, \citealt{Madjarskaetal2004ApJ,Kristaetal2011ApJ}), as well as reported wide-spread magnetic field emergence and cancelation events. Our observations of dynamically evolving bright plasma patches and the formation of spire-like structures following rupturing events are speculated as the occurrence of such reconnections, which leave in their wake newly opened and closed magnetic fields ({\it e.g.}, \citealt{Pariatetal2015AA}). Spire-like structures supportive of an evaporating plasma flow ({\it e.g.}, see 14:38 UT of Figure~\ref{fig:SubStrSeq_E2}), possibly indicate injections of heated plasma into the QS transient feature, likely along newly opened magnetic field lines ({\it e.g.}, \citealt{Shibataetal1997ASPC,Shimojoetal2001ApJ,Pariatetal2015AA}). Note, reported spire structures with visibly bright plasma distributions not decreasing away from the CP site (see 14:30 UT of Figure~\ref{fig:SubStrSeq_E2}), and correlations with evolving bright plasma patches (see 14:10 UT of Figure~\ref{fig:SubStrSeq_E1}), are inconsistent with such interpretations. We emphasize here, additionally the QS transient feature's radiative characteristics (Figures~\ref{fig:SampleImageAlign_CoronalChromo_Evaporation} and \ref{fig:Disschematic}) starkly contrasts such descriptions, {\it i.e.}, those most common of ``classical" jets ({\it e.g.}, \citealt{Allenetal1997SoPh,Shibataetal1997ASPC,Chandrashekharetal2014AA}).

Similarities of our observations to the chromospheric jet evolution detailed by \cite{Liuetal2011ApJ}, mainly, rupturing loops followed by two evolving bright plasma patches ({\it e.g.}, 14:10 UT\,--\,14:12 UT in Figure~\ref{fig:SubStrSeq_E1}), favors the fan-spine magnetic topology typical of jets. Particularly, the outer spine region (see Figure~1 of \citealt{Liuetal2011ApJ}), where continuous injections of heated plasma occur along newly reconnected field lines, and leads to the formation of a diffuse bundle of bright emission \citep{Liuetal2011ApJ,Pariatetal2015AA}, {e.g.}, see Figure~\ref{fig:Disschematic}. Observed radiative emission enhancements near the conclusion of each heating episode, localized to regions out of the direct line of the diffuse bundle of bright 171\,{\AA} emission, are additionally considered here as supportive to a magnetic environment typical of jets. In relation to the fan-spine topology in 3D space, such events could indicate the dome structure as streamlines of falling material \citep{Liuetal2011ApJ}, {\it i.e.}, closed field structures (Figure~\ref{fig:Disschematic}). It is again emphasized, our QS transient maintains its radiative structuring, and visibly bright TR signature over $\approx$\,35\,min time-frame, which as previously pointed out is in stark contrast to that typical of ``classical" eruptive jets, as well as latter evolutions of the chromospheric jet study of \cite{Liuetal2011ApJ}. Thereby, below we postulate to the source of our features unique, and energetically stable radiative nature in a, speculated, magnetic field environment, where ``classical" eruptive jets most commonly occur.

Episodic heating events, witnessed here on $\approx$\,3\,min\,--\,5\,min intervals (Figure~\ref{fig:SubStr_NS-SS-LC}), and the QS transients visibly bright radiative signature over our observational time-frame favors continuous type heating, with moderate reconnection rates ({\it e.g.}, \citealt{Pariatetal2010ApJ,Liuetal2011ApJ,Pariatetal2015AA}). The slower, quasi-steady, reconnection rates and subsequent heating events would explain well the lack of evidence in our observational data for high-speed injections, and significant torsional motion of the TR bundle; consistent with QS expectations \citep{TateArbacheretal2015AAS}. It is speculated such notions point to heated plasma injections, in relation to spire structures with decreasing intensity distributions away from the CP site, resulting from secondary thermodynamic processes \citep{Shibataetal1997ASPC,ShimojoShibata2000AdSpR}
Such spire-like structure formation, which we emphasize includes observed collimated lower ends to ruptured loops, in that respect result from enhanced pressure and temperature gradients ({\it e.g}, \citealt{Shimojoetal2001ApJ,MiyagoshiYokoyama2003ApJ}) at the CP site. We note, as highlighted above, rupturing loop tops that spatially correlate with our QS transient features bundle of bright TR plasma (Figure~\ref{fig:Disschematic}), imply the deposition of additional heated plasma, in the vicinity of the outer spine \citep{Liuetal2011ApJ}. Therefore, both rupturing loop tops, and small-scale evaporation driven jets are speculated to be quasi-steadily supplying our QS transient feature with heated plasma; as such
heating events would explain well its unique radiative signature (Figure~\ref{fig:Disschematic}).

In terms of our QS transient features energetically stable nature, and previous suggestions to a magnetic environment most commonly associated with jets, we speculate below to the lack of a dynamic evolution towards a classical eruptive jet. Recall, as pointed out in $\S$~\ref{sec:intro}, jets most commonly occur in CHs and ARs, large-scale environmental conditions which differ significantly than that of the QS. In particular, CHs and ARs jets align with open coronal magnetic fields, which in terms of ARs is consistent with the large spatial extent of closed field magnetic flux tubes therein that theoretically can be assumed as ``open" \citep{ShimojoShibata2000AdSpR,Schmiederetal2013AA}; assumptions that possibly break down in QS regions, given the much shorter spatial extents, compared to ARs, of its overlying closed magnetic fields. Moreover, as supported by the works of \cite{Shibataetal1994xspy} and \cite{YokoyamaShibata1996PASJ}, the QS environment possibly favors a fan-spine topology where the outer spine is more highly angled away from the solar normal, $\theta$, based on the general geometry of the overlying coronal fields. \cite{Pariatetal2015AA} noted, for $\theta$\,$>$\,20$^{\circ}$ the outer spine actually connects to side boundary of the fan plane, over the typical top boundary, and resultant jet ejections would hit the closed side boundary. In that respect, speculations to such an atypical fan-spine topology of classical jets for our QS transient feature, would explain well the smearing of heated plasma, more parallel to the solar surface, and the radiative structuring inconsistent with expectations \citep{Shibataetal1994xspy,YokoyamaShibata1996PASJ,Orangeetal2013SoPh272O}.

We hypothesize, a coupling of such a fan-spine topology with the general magnetic environment of our QS transient feature possibly acted to inhibit classical jet formation, in particular, the instability build up required to initiate a ``breakout," fast reconnection, phase ({\it e.g.}, \citealt{Hudson2000ApJ,Pariatetal2009ApJ,Pariatetal2015AA}). First, the distributed geometry of the fan-spine, mainly the disjoint outer spine location and angle, could suppress the transition of the 3D null point into an extended current sheet that would be required for the formation of a straight jet \citep{Pariatetal2015AA}. Negligible results for strong magnetic emergence events, in direct relation to our QS transient feature, and torsional radiative motions, further support such a jet inhibiting environment, as they are not indicative of a magnetic instability buildup, such as helicity injections ({\it e.g.}, \citealt{Schmiederetal1995SoPh,Canfieldetal1996ApJ,Pariatetal2015AA}). Moreover, our evidence pointing to the NP, SP, and CP sites as the potential predecessor of episodic heating to the QS transient feature (Figure~\ref{fig:MC_for_Event}), could point to a draining away of their magnetic system's free energy through quasi-steady reconnections as plasma heating.

\section{Conclusion}

Using AIA and HMI observations, recorded on 18 October 2011, during 14:05 UT\,--\,14:40 UT, a QS transient feature, energetically isolated in the TR that shared a common footpoint with two heated coronal loop arcades, and its LOS underlying magnetic field was studied. In radiative images, over our $\approx$\,35\,min observational time domain, the QS transient feature was composed of a diffuse bright bundle of TR heated plasma distributed away from the common footpoint shared with the heated loop arcades (Figure~\ref{fig:Disschematic}). The shared common footpoint site of our QS transient and two heated loop arcades, {\it i.e.}, CP region, was radiatively bright from the chromosphere through the corona during our observations, as well. Radiative image investigations, indicated our QS transient was comprised of magnetically open and closed field structures (Figure~\ref{fig:RDImage_Seq}), consistent with coronal magnetic field extrapolations (Figures~\ref{fig:CMSIntro_Image} and \ref{fig:Disschematic}), where episodic heating events resulted in dynamic evolutions of their radiative emissions (Figures~\ref{fig:SubStrSeq_E1} and \ref{fig:SubStrSeq_E2}). A general heating episode trend, derived from radiative and LOS magnetogram analyses, was established (Table~\ref{tbl:EventMileSt}), and explicitly detailed in $\S$~\ref{sec:Results}.


Observational evidence, {\it e.g.}, loop rupturing (Figure~\ref{fig:SubStrSeq_E1}), spire-like structures (Figure~\ref{fig:SubStrSeq_E2}), enhanced emission of the QS transient's opposing footpoints (Figure~\ref{fig:NS_Spires}), and the embedded dipole of the underlying magnetic field (Figure~\ref{fig:Disschematic}), favored a magnetic topology typical of jets {\it i.e.}, fan-spine topology ({\it e.g.}, \citealt{LauFinn1990ApJ,Pariatetal2010ApJ,Liuetal2011ApJ,Pariatetal2015AA}). It was considered, quasi-steady interchange reconnection events and thermodynamically driven small-scale injections were the source of our QS transient's heated plasma supply, and unique radiative structuring, {\it i.e.}, a diffuse bright bundle of TR plasma (outer spine region; \citealt{Liuetal2011ApJ}), distributed away from a single heated footpoint, where enhanced pressure and temperature gradients likely existed ({\it e.g.}, \citealt{Archontisetal2010AA,Orangeetal2013ApJ,Chandrashekharetal2014AA}). We emphasize, quasi-steady QS transient heating events, $\approx$\,3\,min\,--\,5\,min, and negligible observational evidence for high-speed ejections and torsional motions, favor moderate reconnection rates and evaporating plasma driven by secondary thermodynamic processes, respectively ({\it e.g.}, \citealt{Pariatetal2010ApJ,Shibataetal1997ASPC}, respectively).

Though observations indicated magnetic topologies typical of large-scale eruptive phenomena, {\it e.g.}, classical jets ({\it e.g.}, \citealt{Shibataetal1994xspy,Canfieldetal1996ApJ,YokoyamaShibata1996PASJ,Shimojoetal2001ApJ}), our QS transient maintained an atypical radiative structuring characteristic of such event-types. As such, discussions were presented suggesting that either or both the QS conditions, or generally ``uneventful" nature of the underlying magnetic photosphere possibly inhibited such a dynamic evolution. It was noted, the QS's general magnetic field geometry could have disrupted the typical fan-spine topology, where the outer spine was instead attached to a side boundary and characterized by large degree of separation from the solar surface normal \citep{Pariatetal2015AA}; thus suppressing a catastrophic eruption where the material bundle evolves into a collimated jet \citep{Liuetal2011ApJ}. Additionally, in relation to the general QS conditions, we point out our transient's diffuse bright TR bundle, reflective of a ``smearing" of heated plasma from a bright central core (Figure~\ref{fig:Disschematic}), is consistent with such QS field geometry interpretations. That is, previous works, {\it e.g.}, \citealt{Shibataetal1994xspy,YokoyamaShibata1996PASJ,Orangeetal2013SoPh272O}, have described similar phenomena to result from a current sheet that is highly angled from the solar surface normal, {\it i.e.}, more horizontally directed, as would be expected in our previously described scenario. We also speculated that the LOS magnetic photosphere's evolution, and general strength was possibly unable to provide sufficient instabilities, such as through magnetic helicity injections, required to initiate a ``breakout" phase ({\it e.g.}, \citealt{Pariatetal2009ApJ,Rachmeleretal2010ApJ,Liuetal2011ApJ,Pariatetal2015AA}). This was based on our observational evidence which favored quasi-steady reconnections that likely drained away significant portions of the magnetic system's free energy.

In summary, our work has provided significant insight regarding the environments most commonly associated with large scale solar eruptions, in the presence of the thermodynamic and magnetic conditions prevailing in the QS. Significant evidence presented here supports the QS's capability of sustaining
eruptive topologies, but accompanying energetic build up processes are instead speculated to be resulting in quasi-steady coronal energy deposits over those required for the formation of a ``breakout" or catastrophic eruption, characteristic of CHs and ARs. Observations presented indicating quasi-steady upper TR, and possibly lower corona, interchange reconnections are suggestive to the temperatures
where such magnetic to thermal energy conversions are occurring. In that respect, our work supports an elevated TR role in solar atmospheric energy redistribution processes, and possible solar wind mass feeding, {\it via} localized quasi-steady coronal heating events.
Finally, this work highlights the potential of future cooler atmospheric studies, in particular the TR, considered the heights at which the solar wind originates ({\it e.g.}, \citealt{Parker1963ApJS,Tuetal2005Sci,Liuetal2011AGUFM}), where energetically isolated closed field structures have routinely been witnessed \citep{Orangeetal2013ApJ,Orangeetal2013SoPh272O,Orange2014PhDT}.

\section{Acknowledgements}

The authors greatly appreciate the reviewer's constructive comments on the manuscript. This research was supported by National Aeronautics and Space Administration (NASA) grant NNX-07AT01G and National Science Foundation (NSF) grant AST-0736479. N. B. Orange was also supported by the Florida Space Grant Consortium, a NASA sponsored program administered by the University of Central Florida, grant NNX-10AM01H, as well as Christy and Arlene Orange. N.B. Orange also thanks the University of the Virgin Islands. Any opinions, findings and conclusions or recommendations expressed in this material are those of the author(s) and do not necessarily reflect the views of the NSF or NASA.\\~\\

\bibliographystyle{apj}
\bibliography{qstrjet}

\end{document}